\documentclass{article}
\usepackage{amssymb}
\usepackage{fleqn}
\usepackage{epsfig}
\usepackage{graphicx}
\usepackage{amsmath}
\begin{document}

\begin{center}
\textbf{Radiative and Semileptonic B Meson Decay spectra} \\[0pt]


\vspace*{1cm}

 \textbf{Giulia Ricciardi}\footnote{e-mail address:
Giulia.Ricciardi@na.infn.it} \\ [0pt]

\vspace{0.3cm}
Dipartimento di Scienze Fisiche,\\
Universit\`a di Napoli ``Federico II'' \\
and I.N.F.N., Sezione di Napoli, Italy. \\ [1pt]

$~~~$ \\[0pt]
\vspace*{2cm} \textbf{Abstract} \\[0pt]
\end{center}

We review semi-inclusive charmless B decays, focussing on
threshold logarithmic resumming and on universality of QCD
dynamics in radiative and semileptonic decays.

\vspace*{1cm} \noindent 

\vfill\eject

\setcounter{page}{1} \pagestyle{plain}

\section{Semileptonic  B decays}

 The precise determination of the
magnitude of  the Cabibbo-Kobayashi-Maskawa (CKM) element $V_{ub}$
with a clear uncertainty remains one of the key goals of the heavy
flavor physics programs, both experimentally and theoretically.
The smallest element in the CKM mixing matrix $|V_{ub}|$  plays a
crucial role in the examination of the unitarity constraints and
of the related fundamental questions.

The charmless semi­leptonic $ \bar{B}\rightarrow X_u l \bar{\nu}$
decay channel provides a possible path for the determination of
$|V_{ub}|$. Semileptonic  $B$ decays present several advantages,
such as the possibility of using the systematic framework provided
by the Heavy Quark Effective Theory (HQET), with the additional
assumption of quark-hadron duality. HQET is implemented through
the operator product expansion (OPE) in the form of a heavy quark
expansion \cite{HQExp}. It allows to evaluate inclusive transition
rates as an expansion in inverse powers of the heavy quark mass.
Note that quark hadron duality is not derived from first
principles, but it is a necessary assumptions for many
applications of QCD. For semi-leptonic decays the property of
duality is often referred as global. For non-leptonic decays,
where the total hadronic mass is fixed, it is only the the fact
that one sums over many hadronic states that provides an averaging
(so called local duality). The success of the QCD predictions for
the inclusive semi-hadronic $\tau \rightarrow \nu_\tau + $ hadrons
decays widths is a strong test of global duality \cite{tau_dec}.

 Theoretically, issues regarding the
calculation of the total semileptonic partial width
$\Gamma(\bar{B} \rightarrow X_u l \bar{\nu})$ via OPE are well
understood \cite{OPE, shape2}. The semi-leptonic decay rate can be
calculated as

\begin{equation}
\Gamma=\Gamma_0+ \frac{1}{m_b^2}\Gamma_2 + \frac{1}{m_b^3}
\Gamma_3 + \cdots \label{HQE}
\end{equation}

Here the OPE is both a nonperturbative power series in $1/m_b$ and
a perturbative expansion in $\alpha_s$. The leading term
$\Gamma_0$ is the decay rate of an on-shell b quark treated within
renormalization group improved perturbative QCD. The perturbative
corrections are known to order $\alpha^2$ in the strong
interactions \cite{PT2}. The most remarkable feature of Eq.
(\ref{HQE}) is the absence of a contribution of order $1/m_b$.
That means that non-perturbative corrections are suppressed by at
least two powers of the heavy quark mass. They can be expressed as
matrix elements of higher dimension operators in HQET and
parameterized by
  non perturbative parameters.
 The absence of
contribution of order $1/m_b$ was observed in \cite{abs1sumb} and
it is often referred as Luke's theorem.

Theoretically the total inclusive rate would allow determination
of $|V_{ub}|$ to better than 10\%, the main sources of
uncertainties being the uncertainty in the b quark mass and
uncertainty on potential violation of the underlying assumption of
global quark--hadron duality. However, experimentally, the much
more copious $ \bar{B} \rightarrow X_c l \bar{\nu}$ process, which
has a rate about 60 times higher, does not makes feasible a
measurement over the full phase space.

To overcome this background, inclusive $ \bar{B} \rightarrow X_u l
\bar{\nu}$ measurements utilize restricted regions of phase space
in which the $ \bar{B} \rightarrow X_c l \bar{\nu}$ process is
kinematically highly suppressed. The background is forbidden in
the regions of large charged lepton energy $E_l >  (M^2_B -M^2_D)/
2M_B$ (the endpoint), low hadronic mass $M_X < M_D$ and large
dilepton mass $q^2 > (M_B - M_D)^2$. Extraction of $|V_{ub}|$ from
such a measurement requires knowledge of the fraction of the total
$ \bar{B} \rightarrow X_u l \bar{\nu}$ that lies within the
utilized region of phase space, which complicates the theoretical
issues considerably.

Let us consider the first two kinematic regions for which the
charm background is absent, that is  the large lepton energy
region, $E_l >  (M^2_B -M^2_D)/ 2M_B$, and the small hadronic
invariant mass region $M_X < M_D$. In both cases one needs to
consider the following kinematic region
\begin{equation}
E_X \sim m_b, \quad m_X^2 \sim \Lambda_{QCD} m_b \ll m_b.
\label{kinreg}
\end{equation}
 This kinematic region has sufficient phase space
for many different resonances to be produced in the final state,
so an inclusive description of the decays is still appropriate.
However, in this region, the differential rate is very sensitive
to the details of the wave function of the $b$ quark in the $B$
meson.  The parton level differential distribution at the
end-point region has its own problems, as well, related to the
presence  of large logarithms which spoil  the perturbative
expansion.

A third way to isolate the charmless semileptpnic signal is to use
a selection based on the $q^2$ of the leptonic system. Restriction
of phase space to regions of large $q^2$ also restores the
validity of the OPE \cite{shape6} and suppresses effects due to
the details of the wave function of the $b$ quark in the $B$
meson. Taking only the region kinematically forbidden to $b
\rightarrow c l \bar{\nu}$, $q^2>  (m_B-m_D)^2$ unfortunately
introduces a low mass scale \cite{shape7} into the OPE and the
uncertainties blow up to be of order ($\Lambda_{QCD}/m_c^3$).
Another drawback of this method is the elimination of higher
energy hadronic final states, which may exacerbate duality
concerns.

\subsection{Fermi motion and shape function}

Let us begin by discussing Fermi motion. This phenomenon,
originally discovered in nuclear physics, is classically described
as a small oscillatory motion of the heavy quark inside the
hadron, due to the interaction with the valence quark; in the
quantum theory it is also the  virtuality of the heavy flavor that
matters. Generally, as the mass of the heavy flavor becomes large,
we expect that the heavy particle decouples from the light degrees
of freedom and becomes ``frozen'' with respect to strong
interactions. That is the basic assumption of HQET. However, even
if Fermi motion can be neglected in the ``bulk'' of the phase
space of the decay products, it still plays a role close to
kinematical boundaries, such as the region of interest.
Kinematically, that is easy to show, since a small virtuality of
the heavy flavor in the initial state produces relatively large
variations of the fragmentation mass in the final state.
 A $b$ quark in
a $B$ meson has momentum
\begin{equation}
p_b^\mu= m_b v^\mu+k^\mu
\end{equation}
where $v^\mu$ is the four--velocity of the quark, which we can
take at rest without any loss of generality: $v^{\mu }=(1;0,0,0)$.
$k^\mu = p_b^\mu-m_b v^\mu$ is the residual $b$ quark momentum
after the ``mechanical'' portion of momentum is subtracted off and
it is of order $\Lambda_{QCD}$. If the momentum transfer to the
final state leptons is $q$, the momentum and the invariant mass of
the final state hadrons are
\begin{equation}
p_X^\mu= m_b v^\mu+k^\mu -q^\mu \qquad m^2_X=p_X^2
\end{equation}
The boundary kinematic region is characterized by relations
(\ref{kinreg});  being $k^\mu$ of order $\Lambda_{QCD}$, large
values of $E_X$ can originate only from $m_b v^\mu -q^\mu$,
inducing an almost light--like behavior
\begin{eqnarray}
m_b v^\mu -q^\mu &=& (E_X,0,0,E_X) + O(\Lambda_{QCD}) \nonumber \\
(m_b v -q)^2 &=& O(E_X \Lambda_{QCD}).
\end{eqnarray}
 The invariant mass of the final state hadrons is
\begin{eqnarray}
 m^2_X &=& (m_b v + k - q)^2 = (m_b v - q)^2
 + 2k \cdot (m_b v - q) + O(\Lambda_{QCD}^2) \nonumber \\
 & \simeq & (m_b v - q)^2
 + 2E_X \, k_+
 \label{mx1}
\end{eqnarray}
where  $n^\mu\equiv(1,0,0,1)$ and $k_+=k \cdot n$.
  Over most of phase space, the second
term is suppressed relative to the first by one power of
$\Lambda_{QCD}/m_b$, and so it may be treated as a perturbation.
This corresponds to the usual OPE. However, in the region of
interest the first two terms are of the same order.

This can be also seen in a more compact way, imposing directly
\begin{equation}
 m_b
v^\mu -q^\mu \equiv E_X n^\mu + {k^\prime}^\mu,
\end{equation}
where ${k^\mu}^\prime$ is of order $\Lambda_{QCD}$. Then Eq.
(\ref{mx1}) can be written as
\begin{eqnarray}
 m^2_X &=& (m_b v + k - q)^2 = (m_b v - q)^2
 + 2k \cdot (m_b v - q) + O(\Lambda_{QCD}^2)  \nonumber \\
 &\simeq& E_X k^\prime \cdot n + 2 E_X k \cdot n=
 E_X (k^\prime_+ + k_+).
 \label{mx2}
 \end{eqnarray}

A fluctuation in the heavy quark momentum of order $\Lambda
_{QCD}$ in the initial state produces a variation of the final
invariant mass of the hard subprocess of order
\begin{equation}
\delta m_{X}^{2}\sim O\left( \Lambda _{QCD}\,E_{X}\right) .
\label{ampl}
\end{equation}
An amplification by a factor $E_{X}$ has occurred, as anticipated.

The differential rate in this  region is therefore sensitive  to
the wave function $f(k_+)$ which describes the distribution of the
light cone component of the residual momentum of the $b$ quark.
The shape function is a non-perturbative function and cannot be
calculated analytically, so the rate in that region is model
dependent even at leading order in $\Lambda _{QCD}/m_b$. If we
consider a heavy quark with the given off-shell momentum  and a
final state consisting of a massless on-shell quark at the tree
level, we find for the shape function
\begin{equation}
f(k_{+})^{part}=\delta \left( k_{+}+\frac{(m_b
v-q)^2}{2E_{X}}\right) +O\left( \alpha _{S}\right), \label{basres}
\end{equation}
as it should be, since  $m_X^2=0$. Selecting the hadronic final
state, i.e. $k_{+}$, we select the light-cone virtuality of the
heavy flavor which participates in the decay.

We note that even with the amplification effect, Fermi motion
effects are irrelevant in most of the phase space, where typical
values for the final hadron mass are
\begin{equation}
m_{X}^{2}\sim O(E_{X}^{2}),  \label{bulk}
\end{equation}
 in agreement with physical intuition.

 The shape function $f(k_+)$ is
written as a function of  $k_+=k^0 + k_\|$; spatial components
$k_\|$ and $k_\bot$ have been defined with respect to the $m_b
v^\mu-q^\mu$ direction (roughly the recoiling u quark). At this
order, possible contributions  due to $k_\bot$ are ignored. The
shape function
 can also be seen as a resummation of the
OPE to all orders in $E_X \Lambda_{QCD}/m^2_X$
\cite{shape2,shape1,shape3}.

Because the shape function depends only on parameters of the B
meson, it is reasonable to expect that this leading order
description holds for any B decay to a light quark.  In general,
we aspect QCD to factorize long distance effects into structure
functions, with universal characteristics. In particular, one
would like to get an estimation of structure functions for the
semileptonic decays through the differential rate of $B
\rightarrow X_s \gamma$ decay. However, due to the different
kinematics between two body and three body processes, and to the
presence of more than one energy scale ($E_X$, $m_X$ and $m_b$),
universality of structure functions is not trivially applicable
\cite{noi1,noi2,noi3}.

\subsection{Perturbative resumming of large logarithms}

In general, the differential partial width is given by the
convolution of  a non-perturbative  structure function with the
perturbative calculable parton level differential distribution.
Large remnants of the long distance dynamics occur also at the
perturbative level by the presence of large logarithms near the
threshold regions of phase space. Threshold resummation is a well
known calculation technique which organizes the logarithmic
enhancements to all orders in perturbation theory, thereby
extending the QCD predictive power. In perturbative QCD, the
hadronic subprocess in $\bar{B} \rightarrow X_u l \bar{\nu}$
consists of a heavy quark decaying into a light quark which
evolves later into a jet of soft and collinear partons. The series
of large infrared logarithms is due to an only partial
cancellation of infrared divergencies in real and virtual
diagrams. Let us consider f.i. the light quark produced in a
process with a hard scale $Q$, evolving into a jet whose invariant
mass is kinematically limited to a value $m$ well below $Q$; the
smaller integration region of the real diagrams induces a
left-over double logarithm in the ratio $Q/m$. Multiple gluon
emission occurs at high orders of perturbation theory, originating
a double logarithmic series \cite{sterman,cattren1}.

\section{Radiative B decays}

Let us  consider  the  radiative decay with a real photon in the
final state,
\begin{equation}
\label{bsgamma} B \, \rightarrow \, X_s \, + \, \gamma
\end{equation}
A systematic resummation is best done in the $N$-moment space or
Mellin space, because it leads to the exponentiation of the
logarithmic corrections. In the $N$-moment space the threshold
region corresponds to the limit $N\rightarrow\infty$
\cite{catani}. Let us consider the Mellin transform of the
normalized spectrum
\begin{equation}
\label{mellindist1} \Gamma_{R,N} \equiv \int_0^1 (1-t_s)^{N-1}
\frac{1}{\Gamma_R}\frac{d\Gamma_R}{d t_s} \, dt_s
\end{equation}
In  Mellin space threshold logarithms manifest themselves as a
series in $ \log N$.  The order by order perturbative expansion
contains double logarithmic contributions and it has the following
schematic form
\begin{equation}
\Gamma_{R,N} = 1+\,\alpha \,  ( c_{12} \,L^{2}+c_{11}\,L+c_{10})
\,+ \,\alpha^{2}\,
(c_{24}\,L^{4}+c_{23}\,L^{3}+c_{22}\,L^{2}+c_{21}\,L
\,+c_{20}\,)+\cdots    \label{exp}
\end{equation}
where $L \equiv \log N$ and $\alpha $ is the running coupling
evaluated at the hard scale $m_b$.

The logarithmic terms have an exponential structure and we can
write the following factorized form
\begin{equation}
\label{mellindist2} \Gamma_{R,N} \,=\, C_R(\alpha)\,
\sigma_N(\alpha) \, + \, d_{R,N}(\alpha),
\end{equation}
 where the form factor $\sigma_N$ in $N$-space contains all the large
logarithms to all orders in $\alpha$
\begin{eqnarray}
\label{thresum} \sigma_N(\alpha) \, &=& \, e^{ G_N(\alpha) }
 =\exp \left[ \sum_{n=1}^{\infty
}\sum_{k=1}^{n+1}G_{n\,m}\,\alpha^{n}\,L^{k}\right] = \notag \\
&=&\exp \left[ G_{12}\alpha\,L^{2}+G_{11}\alpha
\,L+G_{23}\alpha^{2}\,L^{3}+G_{22}\alpha^{2}\,L^{2}+
G_{21}\alpha^{2}\,L+G_{34}\alpha ^{3}\,L^{4}+\cdots \right].
\end{eqnarray}
The coefficient function $C_R(\alpha)$  has a perturbative
expansion of the form
\begin{equation} C_R(\alpha) \, = \, 1 \, + \, \alpha \,
C_R^{(1)} \, + \, \alpha^2 \, C_R^{(2)} \, + \, O(\alpha^3),
\end{equation}
where $C_R^{(i)}$ are numerical coefficients, independent on $N$.
 The
remainder function has a similar perturbative expansion; it
 depends on N, but it tends to zero in the threshold
limit
\begin{equation}
d_{R,N}(\alpha) \, \rightarrow \, 0 \qquad {\rm
for}~N\,\rightarrow\,\infty.
\end{equation}

The double sum in the exponent  (\ref{thresum}) is usually
organized as a series of functions $g_i$
\begin{equation}
\label{series}
 \sigma_{N}\left( \alpha \right) =\exp \left[
L\,g_{1}\left( \beta _{0}\alpha L\right) +\sum_{n=0}^{\infty
}\alpha^{n}\,g_{n+2}\left( \beta _{0}\alpha L\right) \right] =\exp
\left[ L\,g_{1}\left( \lambda\right) +\,g_{2}( \lambda) +\alpha
\,g_{3}( \lambda) +\alpha^{2}\,g_{4}( \lambda) +\cdots \right].
\end{equation}
The functions $g_{i}\left( \lambda \right) $ have a power-series
expansion:
\begin{equation}
g_{i}\left( \lambda \right) =\sum_{n=1}^{\infty }g_{in}\lambda
^{n}.
\end{equation}
where $\lambda\equiv \beta_0 \,\alpha\, L$ and $\beta _{0}$ is the
first coefficient of the $\beta$ function expansion in $\alpha$.
By mantaining in the exponent of $\sigma_N$ only the function
$g_1$ one approximates at leading order (LO); similarly keeping
$g_1$ and $g_2$ one approximates at next-to-leading order (NLO),
 $g_1$, $g_2$ and $g_3$ at
 next-to-next-to leading order (NNLO) and so on.
 This is equivalent to resumming series up to
$L (\alpha L)^n$, $(\alpha L)^n$ and $%
\alpha (\alpha L)^n$, and so on.
 The explicit expressions of $g_i$ are
known  up to NNLO \cite{akhouri,ugosol,ugnnlo,noi1}.

The exponent in (\ref{thresum}) can be obtained by means of the
following resummation formula \cite{sterman,cattren1,cattren2}
\begin{equation}
G_N(\alpha) = \int_0^1 dz \frac{ z^{N-1} - 1 }{1-z} \Bigg\{
\int_{Q^2(1-z)^2}^{Q^2(1-z)} \frac{dk_t^2}{k_t^2}
A\left[\alpha(k_t^2)\right] \, +  B\left[\alpha(Q^2(1-z))\right]
 +  D\left[\alpha(Q^2(1-z)^2)\right]
\!\!\Bigg\}, \label{expresum}
\end{equation}
where $Q=2 E_X$ is the hard scale of the process: in radiative
decays, $Q=m_b$. The functions $A\left(\alpha\right)$,
$B\left(\alpha\right)$ and $D\left(\alpha\right)$ have a standard
fixed order expansion in $\alpha$, with numerical coefficients
\begin{equation}
 A\left(\alpha\right) =
  A_{1}\alpha+A_{2}\alpha^{2}+\cdots, \quad
B\left( \alpha\right) = B_1\alpha+B_2\alpha^2+\cdots,
\quad D\left(\alpha\right) =D_1\alpha +D_2\alpha^2+\cdots .
\end{equation}
$A\left(\alpha\right)$ describes the emission of partons which are
both soft and collinear, $B\left(\alpha\right)$ describes hard and
collinear partons, while $D\left(\alpha\right)$ describe partons
which are soft and at large angles. The values of the known first
coefficients of the functions $A\left(\alpha\right)$,
$B\left(\alpha\right)$ and $D\left(\alpha\right)$ are reported in
\cite{noi1}.

By truncating $\alpha $ expansions for the functions $A(\alpha)$,
$B(\alpha)$ and $D(\alpha)$ in eq.~(\ref{expresum}) one is
implicitly assuming $\alpha\ll 1$. That is not always correct
since the running coupling $\alpha\left(k_t^2\right)$ is
integrated over all gluon transverse momenta $k_t$ from the hard
scale $Q$ down to zero: inside the integration region, the Landau
pole is hit and the running coupling diverges. A prescription has
to be assigned to give a meaning to the formal expression
(\ref{expresum}). Even after integration,  an effect of the
presence of the Landau pole persists:  the series in
eq.~(\ref{series}) is divergent as the higher order functions have
factorially growing coefficients \cite{ugnnlo,gardi}. By
truncating it to its first few terms, we stay within the so called
fixed logarithmic accuracy. Moreover, the functions $g_i(\lambda)$
have branch cuts starting at $\lambda=1/2$  and going up to
infinity. When $ \lambda \rightarrow \frac{1}{2}^{-}$, by
definition of $\lambda$, a singularity in $N$-space occurs  at $
N\rightarrow \exp \left[ 1/2\beta _{0}\,\alpha _{S}\left(
Q^{2}\right) )\right] $, that is at $N_{crit}\sim Q/\Lambda$
($\Lambda$ is the QCD scale)\footnote{ Let us also observe that
the degree of singularity of the functions $g_{i}$ for $\lambda
\rightarrow 1/2,$ and therefore also of the form factor, increases
with the order of the function, i.e. with $i$  \cite {gardi}.}.
The form factors are then formally well defined up to a critical
value $N_{crit}$, above which they acquire a (completely
unphysical) imaginary part.

 To return from $N$ space  to the
physical space one uses the Mellin inverse transform:
\begin{equation}
\label{inverse} \sigma(t_s; \, \alpha) \, = \,
\int_{C-i\infty}^{C+i\infty} \frac{dN}{2\pi i} (1-t_s)^{-N} \,
\sigma_N(\alpha).
\end{equation}
 The evaluation of the inverse transform requires a prescription,
in order to overcome the problems just mentioned and to obtain a
form factor formally well defined in the whole $t_s$-space, which
resums all the logarithmic terms at the requested order. Several
strategies can be pursued towards this result. The simplest
possibility is to restrict oneself to a fiducial region in
$N$-space below $N_{crit}$. Another possibility
 is to use the
so-called minimal-prescription \cite{minimalpre}, which
regularizes
 the form factor by means of an additional
prescription for the contour integration in $N$-space in
(\ref{inverse}).
  The problem of the presence of an integration over the Landau pole in resummation
  formulas,
and of the ambiguities associated,  has  been also examined  in
the contest of the occurrence of infrared renormalons, to get
information about non-perturbative effects, in the form of
power-suppressed corrections to the cross sections (see f.i.
\cite{gardi}).

Another approach is to give a prescription for the infrared pole
directly in $N$-space, in such a way that the form factors are
well-defined for any $N$. Then it is not necessary to give a
prescription for the Mellin inverse transform.
 A  recent analysis \cite{pheno06}
substitutes an effective coupling to the standard running coupling
in resumming formula (\ref{expresum}). The effective coupling is
built in a way to mantain the high energy behaviour of the
standard running coupling, without reaching the Landau pole at low
energies. Therefore the effective coupling never straddles far
outside the perturbative domain. The resummation formula is free
of Landau pole pathologies, and no prescription is needed because
$\sigma_N(\alpha)$ is analytic on the integration contour
(\ref{inverse}).

 The effective form factor  may
also include absorptive effects related to the coupling constant
 in order to derive an improved expression for the
resummation formula. Such absorptive effects are related to the
decay of time-like gluons in the jet evolution. As well known from
perturbation theory, at higher orders, one has to consider
multiple emissions off the heavy and the light quark in $B$
 and secondary emissions off the radiated gluons.
Because of the presence of these higher-order terms, the coupling
in the resumming formula is evaluated at the transverse momentum
of the primary emitted gluon \cite{Amati:1980ch}. By including the
$-i\pi$ terms in the integral over the discontinuity, i.e. the
absorptive effects, usually neglected, the coupling in the
resumming formulas is replaced by an effective coupling, evaluated
at the transverse momentum of the primary emitted gluon
\cite{pheno06, model1}
 \begin{equation} \alpha
\, \to \, \tilde{\alpha}(k_{t}^2).
\end{equation}

There is another advantage by using a perturbative approach with
an effective coupling. In the standard approach, after resumming
the large perturbative logarithmic contributions, one has to
postulate  a physically motivated non-perturbative model.
Generally, an {\it ad-hoc} non perturbative form factor is
convoluted with the perturbative distributions. Universal aspects
of QCD dynamics, common to  different processes, are not easily
discovered this way. On the other hand, by describing different
processes with the appropriate perturbative formulas, and  the
same effective coupling by assumption, one can deal simultaneously
with perturbative and non-perturbative effects (without
introducing model-depending parameters),  avoid mistakes related
to double counting, and underline universal effects by comparing
with the data \cite{pheno06, model1, Dokshitzer:1995qm}.

\section{Universal aspects of QCD dynamics}

Let us now consider the decay
\begin{equation}
\label{semilep}
B \, \rightarrow \, X_u \, + \, l \, + \, \nu.
\end{equation}
It is possible to obtain a factorized form for the triple
differential distribution, the most general distribution in
process (\ref{semilep}) (its integration leads to all other
spectra):
\begin{equation}
\label{tripla} \frac{1}{\Gamma}\frac{d^3\Gamma}{dx du dw} \,=\,
C\left[x,w;\alpha(w\,m_b)\right]\,
\sigma\left[u;\alpha(w\,m_b)\right] \, + \,
d\left[x,u,w;\alpha(w\,m_b)\right],
\end{equation}
where:
\begin{equation}
w \,\equiv \, \frac{2 E_X}{m_b}~~~~~~~~(0\le w \le
2),~~~~~~~~~~~~~~~~~~~~~ x \,\equiv \, \frac{2
E_l}{m_b}~~~~~~~(0\le x \le 1)
\end{equation}
and
\begin{equation}
u \, \equiv  \, \frac{E_X - \sqrt{E_X^2 - m_X^2} }{E_X +
\sqrt{E_X^2 - m_X^2} } \, = \, \frac{1 - \sqrt{1 -
\left(2m_X/Q\right)^2} }{1 + \sqrt{1 - \left(2m_X/Q\right)^2} } \,
\simeq \, \left(\frac{m_X}{Q}\right)^2~~~~~~~~~~(0\le u \le 1).
\end{equation}

We have called  $Q$ the hard scale of the process and, at the
threshold,  set $Q= 2 \, E_X$. Both the logarithms and the
argument of the running coupling depend on the kinematics of the
problem, by means of the hard scale $Q$. There are two important
kinematical differences with respect to the radiative case. First,
in the three body semileptonic decay, the distribution also
depends on the charged lepton energy $E_l$. Second,  while in the
radiative decay it is always $Q = 2 \, E_X = m_b$, in process
(\ref{semilep}) there are regions of phase space  where $E_X$ is
substantially reduced \footnote{One can think, f.i., to the
kinematical configuration with a large invariant mass for the
lepton pair}.

In analogy with what done in the previous section, we can study
the distribution in the Mellin space, defining
\begin{equation}
 \sigma_N(\alpha) \, \equiv \, \int_0^1 du \, (1-u)^{N-1}
\, \sigma(u;\,\alpha) \label{Mellin:semi}
\end{equation}
At this level, there is universality among radiative and
semi-leptonic decays, meaning that the same QCD form factor
$\sigma_N$  appears in both distributions (\ref{mellindist2}) and
(\ref{Mellin:semi}). Consequently,  the form factor in the
physical space is the same  $\sigma(u;\,\alpha)$. It is evaluated
at the argument $u \simeq m_X^2/(4 E_X^2)$ in the semileptonic
case. In the radiative case, by imposing the kinematical relation
between hadronic energy $E_{X_s}$ and hadronic mass $m_{X_s}$, we
have
\begin{equation}
u|_{E_{X_s} = m_b/2(1+m_{X_s}^2/m_b^2)} \, = \, t_s.
\end{equation}
where $t_s \equiv m_{X_s}^2/m_b^2 $.

 The coupling constant argument is
set at the hard scale $Q=2 E_X$ in both processes; in the
radiative decay, that implies it is fixed to $m_b$.

This simple connection between radiative and semileptonic
processes is sometimes lost when one passes to the double or
single differential distributions for the processes
(\ref{semilep}) \cite{ugosol,noi1,noi2,noi3}. All double and
single distributions
 are obtained by integrating the triple
differential distribution (\ref{tripla}). As seen before, in
semileptonic decays the hadronic energy $E_X$ is not fixed and it
can be integrated over; this affects the infrared structure of the
distribution, since the form factor $\sigma$ depends on $E_X$
\cite{ugosol,noi1,noi2,noi3}. This class of distributions, f.i.
the single differential distribution in $m_X$, cannot be directly
compared with the radiative spectra. The structure of the
threshold logarithms is not the same and
   there is no universality of long distance
effects. On the other side,  distributions as the single
differential one in $E_X$, where the energy $E_X$ is not
integrated over, keep the infrared structure of the radiative
spectrum. They can directly be related via short-distance
coefficients to the radiative spectrum.


\end{document}